# Activation energy distribution in thermal quenching of exciton and defect-related photoluminescence of InP/ZnS quantum dots

S.S. Savchenko[a], A.S. Vokhmintsev[a], I.A. Weinstein[a,b,*]

[a] NANOTECH Centre, Ural Federal University, 19 Mira str., Ekaterinburg, 620002, Russia

[b] Institute of Metallurgy, Ural Branch of the Russian Academy of Sciences, 101 Amundsena str., Ekaterinburg 620016, Russia

* Corresponding author: i.a.weinstein@urfu.ru (I.A. Weinstein)

**Abstract.** Thermal quenching is one of the essential factors in reducing the efficiency of radiative processes in luminophores of various nature. The emission activity of low dimensional structures is influenced also by multiplicity of parameters that are related to synthesis processes, treatment regimes, etc. In the present work, we have investigated the temperature dependence of photoluminescence caused by exciton and defect-related transitions in ensembles of biocompatible InP/ZnS core/shell nanocrystals with an average size of 2.1 and 2.3 nm. The spread in the positions of energy levels is shown to be due to size distribution of quantum dots in the ensembles under study. For a quantitative analysis of the experimental data, we have proposed a band model accounting for the Gaussian distribution of the thermally activated barriers in the photoluminescence quenching processes. The model offers the thermal escape of an electrons from the core into the shell as the main mechanism for non-radiative decay of excitons. In turn, the quenching of defect-related emission is predominantly brought about through the emptying of the hole capture centers based on dangling phosphorus bonds. We have revealed the correlation between size distributions of quantum dots and scatter of the activation energy of exciton luminescence quenching. The developed approach will give further the possibility to optimize technological regimes and methods for band engineering of indium phosphide-based type-I quantum dots.

# 1. Introduction

Colloidal quantum dots (QDs) belong to a unique class of nanoscale objects whose electronic and optical properties occupy an intermediate position between molecules and three-dimensional bulk materials composed of them [1–3]. Temperature-dependent features of luminescence of II-VI and III-V semiconductor nanocrystals provide their wide applied prospects for designing laser and LED media, composite phosphors, luminescent biomarkers, solar energy concentrators, photovoltaic devices, temperature sensors etc. [4–12]. From the standpoint of optimizing their operational specifications, the range near room temperature and above is decisive. Meanwhile, cryogenic investigations of the optical properties of zero-dimensional objects allow one to make deeper insight into the fundamental regularities of radiative and non-radiative recombination, to describe the structure of electron-hole states, to study the relaxation and transport of excitations, to analyze the mechanisms of blinking, thermal quenching, and many other processes occurring at micro - and nanoscale levels.

As the temperature rises, the intensity of exciton luminescence in most QDs drops. In other words, thermal quenching happens. On the one hand, an increase in the contribution of non-radiative transitions can be caused by internal exciton-phonon interaction processes, thermal activation through the barrier between the potential energy surfaces of the excited and ground states, as well as thermally assisted tunneling [13, 14]. On the other hand, noticeable emission quenching can be a consequence of phenomena associated with external mechanisms, in particular, with capturing excited charge carriers by traps in the region of the surface and various interfaces [14, 15].

At present, the processes of thermal quenching have been studied in sufficient detail for QDs with a core based on II-VI compounds [16–18]. Besides there are a number of independent data on the temperature influence on luminescent properties of indium phosphide that is one of the promising materials for creating environmentally friendly and biocompatible QDs based on III-V compounds. For instance, it is reported in [19] that from room temperature to 800 K, a significant quenching of luminescence occurs in InP nanocrystals with various kinds of surface passivation. In the process, a ZnS shell or small inorganic ligands on the surface are the reason for improving the high-temperature stability of QDs, and electron capture is the main quenching mechanism. The authors of [20, 21] claim that, when heated from cryogenic to room temperature, InP nanocrystals of various sizes demonstrate a decline in the PL intensity over the entire spectral range. Alongside this, a red shift of the exciton PL maximum and a strong quenching of low-energy emission are observed. As a result, at room temperature, the spectrum exhibits a broad peak with an extended tail. The luminescence parameters of InP/ZnS nanocrystals exposed to temperature ranged from 2

K to 510 K were explored in [22, 23]. Changes both in the energy of the maximum and in the halfwidth of the PL band are found to be due to the interaction of excitons with acoustic phonon modes. The authors of [24] made a note of different behavior of PL quenching caused by exciton transitions and optically active centers based on crystal defects. With changing the temperature, a similar shift of the maximum, as well as a strong quenching of low-energy defect-related PL was also observed in [25–27]. In this case, a quantitative analysis of the processes of thermal quenching of exciton emission was performed for a series of InP/ZnS QDs of different sizes (2.0 – 4.7 nm) and in the range of 80–300K, accounting for the exciton-phonon interaction with longitudinal acoustic and optical modes [27].

So we summarize that there are many factors affecting the efficiency of the luminescent response of the core/shell QDs. Firstly, they are directly connected with the phase-chemical composition, intrinsic properties and structural characteristics of the core material. Secondly, the semiconductor shell itself and its features contribute to forming active quenching centers based on various defects in the interface regions. Finally, the temperature behavior of luminescence is influenced by stabilizing molecules and the solvent, with both controlling the relaxation of surface atoms and being responsible for the appearance/disappearance of surface channels of non-radiative relaxation of excited charge carriers. All of the factors mentioned above predetermine the peculiarities of each nanocrystal separately. However, the samples are, as a rule, ensembles combining QDs spread in size, shape, and other structural and physicochemical parameters. Variation in size, for example, governs the position of the energy levels; therefore, the scatter in the energy gap width, the magnitude of exciton and defect-related optical transitions, the height of various energy barriers responsible for the appropriate channels of radiative and non-radiative relaxation of excitations should be expected.

Earlier [28–30], we studied the effects of inhomogeneous broadening of exciton bands in optical absorption spectra of InP/ZnS QDs with different average sizes, as well as the temperature shift of exciton levels due to interaction with effective low-energy vibrations. The present paper is aimed at investigating the mechanisms of thermal quenching of luminescence in colloidal InP/ZnS nanocrystals in the range 6.5–296 K, taking into account possible effects of the size distribution of QDs in the ensemble.

## 2. Methods

In this work, we studied photoluminescence spectra for two ensembles of water-soluble InP/ZnS QDs, which were synthesized by NIIPA (Dubna, Russia) based on the method [31] using safe amino-phosphine-type precursors. An average particle size estimated from the position of the exciton absorption band was 2.1 (QD-1) and 2.3 nm (QD-2). For luminescence measurements in the temperature range 6.5–296 K, the colloidal solutions were drop casted onto 1-mm-thick quartz substrates and then dried at room temperature to form a film. Transmission of the substrates was 94% in the range 300–900 nm and gradually decreased to 85% at 190 nm. Under the experimental conditions, the substrates did not exhibit their own luminescent response. PL emission was dispersed with a Shamrock SR-303i-B spectrograph (Andor, Inc.). The input slit width was 50 μm, and the diffraction grating had a groove density of 150 l/mm. A Newton$^{EM}$ DU970P-BV-602 CCD matrix (Andor, Inc.) cooled to -80 °C was used as a detector. Resolution of the measuring system was 1.3 nm. The spectra were read out in the full vertical binning mode, exposure time was 200 ms, the data obtained were then smoothed using the Savitsky-Golay algorithm. To plot the spectra against the photon energy a spectral correction was performed. Control of the sample temperature was implemented by means of a CCS-100/204N helium closed cycle refrigerator (Janis Research Company, LLC). Measurements were carried out at different temperatures: $T$ = 6.5 K, over the range from 10 to 100 K in 10 K, and from 100 to 296 K with a step of 20 K. The presented results were obtained under UV LED excitation with an intensity maximum at 372 nm and average power density of 5 mW/cm$^2$. PL quantum yield of the nanocrystals was estimated at room temperature within the framework of a well-known procedure [32]. Rhodamine 6G was used as a reference. The values obtained were 27% for QD-1 and 10% for QD-2.

## 3. Results

Fig. 1 shows photoluminescence spectra of the QDs measured at different temperatures and normalized to the maximum intensity at $T$ = 6.5 K. The room temperature luminescence bands have an asymmetric shape characterized by a less steep low-energy part. For QD-1, the PL spectrum has a maximum intensity at $E_x$ = 2.32 eV and a halfwidth $H$ = 296 meV. The larger QD-2 nanocrystals exhibit $E_x$ = 2.13 eV and $H$ = 364 meV. Upon cooling, the intensity increases in both samples, and the $E_x$ position shifts to the blue region towards 2.36 eV for QD-1 and 2.16 eV for QD-2 at 6.5 K. In this case, spectra shape changes significantly, the halfwidth increases to 705 and 559 meV, respectively. The observed broadening is associated with the intense shoulders in the low-energy region of the experimental spectra distinctly revealed at $T$ <160 K for QD-1 and at $T$ <100 K for QD-2. For the smaller nanocrystals this feature turns into the band with the 2.07 eV local maximum at 6.5 K. Such evolution points to the presence of an additional component in the emission of the samples (see also Fig. 2). Integrated luminescence intensity in the range 1.5–2.75 eV changes 13 times for QD-1 and 9 times for QD-2 in the studied temperature interval.

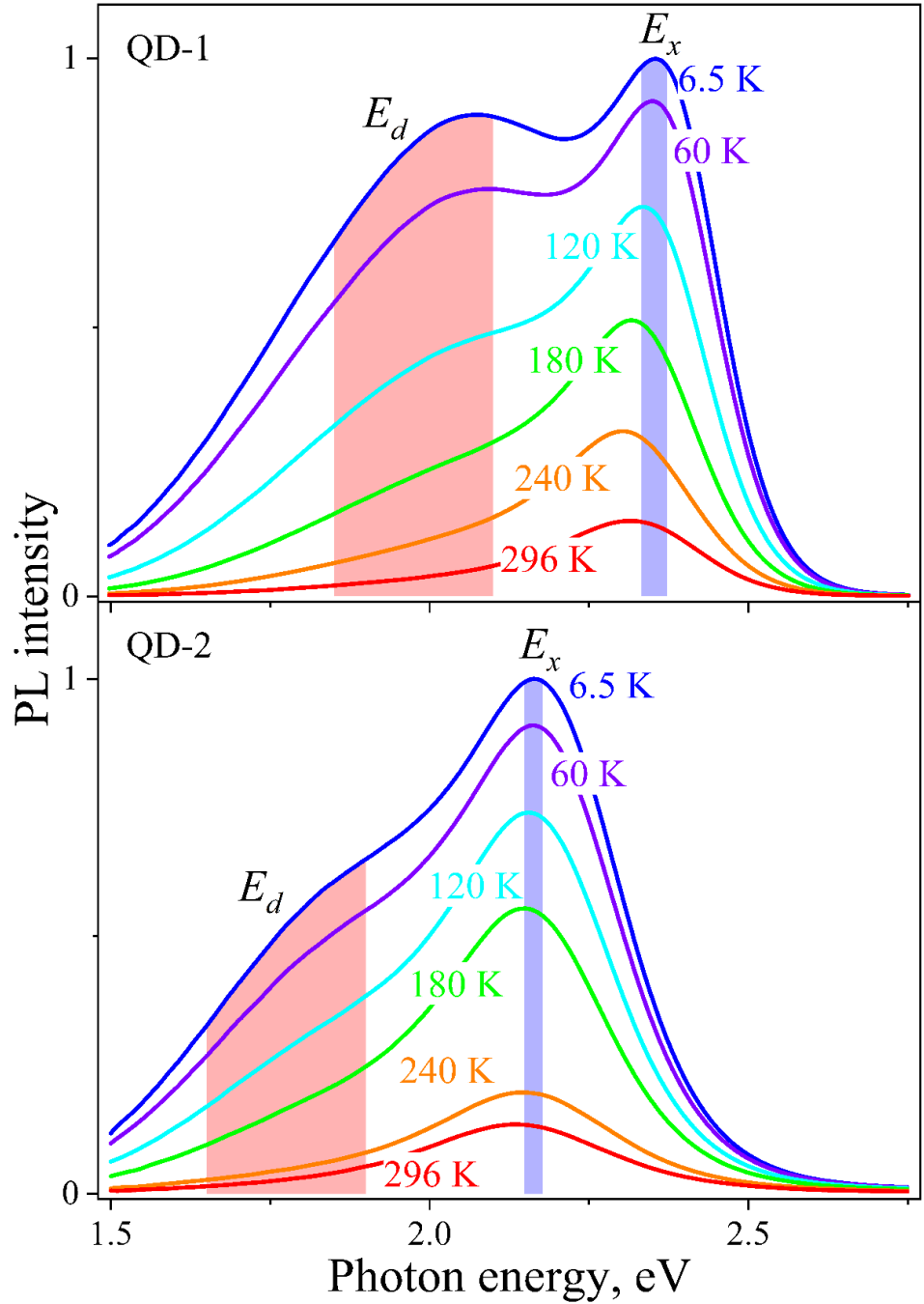

**Figure 1.** Experimental PL spectra for the InP/ZnS QDs. The measurement temperatures are shown next to the corresponding curves. The intensity of all curves is normalized to the maximum value at 6.5 K. For the description of the colored areas, see section 4.4.

## 4. Discussion

*4.1. Temperature evolution of the InP/ZnS photoluminescence*

For clear comparative analysis of the changes in the shape, the measured PL spectra are presented in Fig. 2 after normalization to the appropriate maxima at different temperatures. The position of the $E_d$ maximum for the hidden component responsible for the temperature dynamics in the low-energy part of the spectrum was calculated using the second-order derivative spectroscopy [29] (see Fig. 2, dashed line). The observed transformations of the luminescence spectra in the QDs at hand may be due to the presence of several radiative channels based on excitons, various defect-related centers, etc. Moreover, the mechanisms of PL thermal quenching are also different in origin.

The temperature dependencies of photoluminescence in InP/ZnS nanocrystals of different sizes have been studied earlier in [23–27] (see data in Table 1). These papers are attributed the high and the low energy spectral components to the exciton transitions and the presence of defects, respectively. It should be underscored that, at low temperatures, most of the known photoluminescence spectra contain defect-related emission, $E_d$ (see Table 1). An exception is provided by the samples in [23] because the spectrum lacks a clearly pronounced component, $E_d$. Moreover, the largest QDs exhibit no defect-related emission in the size series of the 2 – 4.7 nm samples [27]. The authors of [26] spot three components in the PL spectrum. The main one corresponds to recombination through the doublet of bright and dark exciton states. The higher energy component is ascribed to another bright state lying above the doublet. The low-energy band matches recombination through deep traps or defect states. Along with exciton emission, photoluminescence of defects in InP/ZnS is also confirmed in [27]. The work [25] reports that a decline in temperature contributes to the appearance of bands shifted relative to the exciton maximum by about 300 meV towards lower energies, with their intensity dropping as the QDs grow in size. In the opinion of the authors, deep traps explain this fact. The current research emphasizes that the emission intensity of $E_d$ also increases with a decrease in the nanocrystal size (see Fig. 2). This probably points to a relationship between the emission and surface states. The energy differences amount to $E_x - E_d$ = 320 meV (for QD-1) and 360 meV (for QD-2), which is quite consistent with the data and conclusions in [25].

Thus, given the foregoing, the $E_x$ band observed is due to exciton emission, and the broad low-energy component $E_d$ may be caused by lattice defects in InP/ZnS. First and foremost, the optical properties are affected by dangling bonds of indium $DB_{In}$ and phosphorus $DB_P$ atoms on the surface of the core, which create donor and acceptor levels in the energy gap of QDs, respectively [20]. Besides, substitutional $Zn_{In}$ defects (zinc atoms in the crystalline positions of indium) can arise [33]. Calculations based on the density-functional theory indicate that the presence of $Zn_{In}$ leads to the formation of defect states 50 – 250 meV above the top of the valence band, depending on the position in the lattice.

We have previously shown in [34, 35] that the shape of the experimental spectra at different temperatures is well reproduced through a numerical deconvolution into several Gaussian components. However, the obtained spectral parameters and their dependencies provide no physically substantiated interpretation of the observable temperature behavior, in particular, within the exciton-phonon interaction. At our glance, the temperature features in the PL spectra can be explained by the scatter of the parameters of quantum dots in the ensemble. For InP/ZnS nanocrystals, the effects of such a spread have been independently examined based on the Kennard-Stepanov relation and on estimating the distribution of the luminescent-state lifetimes at room temperature [36]. We also analyzed these effects for the QD-1 and QD-2 samples within an experimental and theoretical analysis of the halfwidth of optical absorption bands in a wide temperature range [28, 30, 37].

*4.2. Inhomogeneous broadening of the exciton band*

The method of estimating the halfwidth of the exciton band $H_x = 231 \pm 30$ meV for QD-1 and $315 \pm 30$ meV for QD-2 is illustrated in Fig. 2. It is worth noting that the $H_x$ magnitudes remain unchanged within an experimental error in the entire temperature range and do not exceed the corresponding values of 290 and 370 meV for the halfwidth of the first exciton band in the optical absorption spectra at the same samples [30].

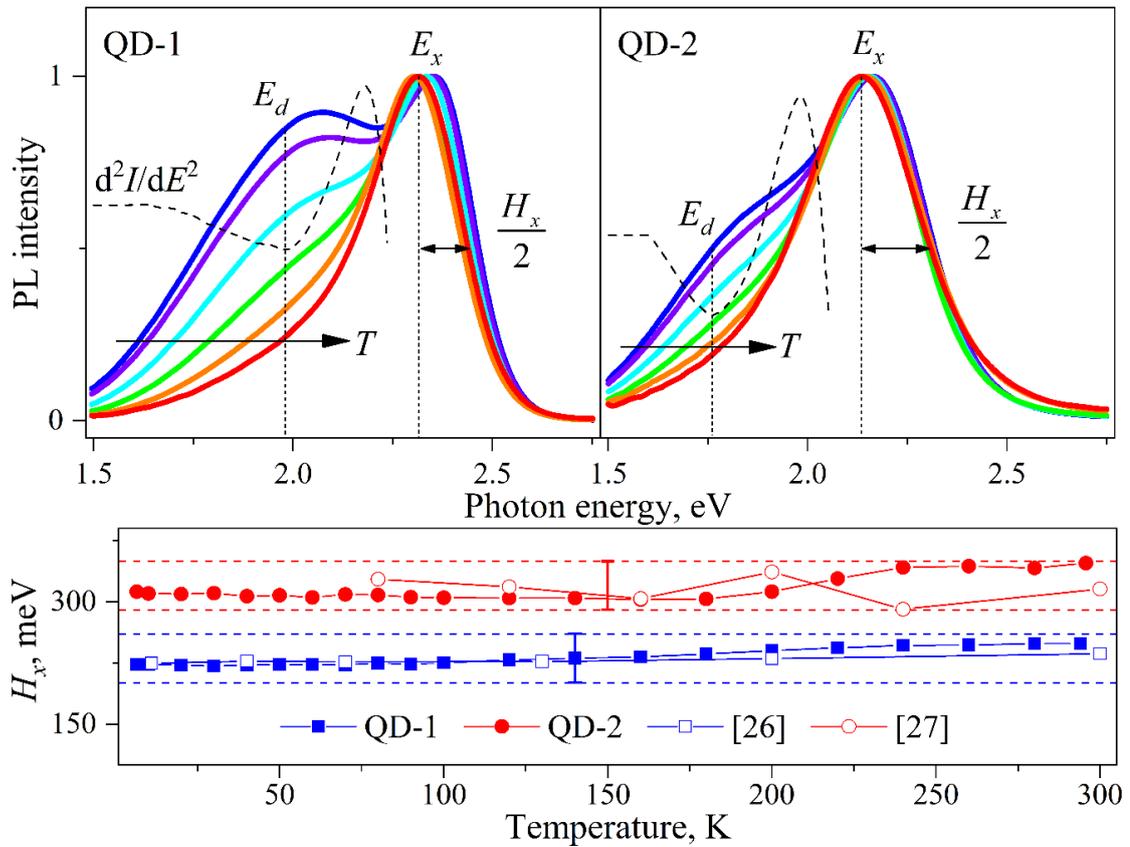

**Figure 2.** The shape of the PL spectra analyzed for the samples tested.
The dashed line indicates the spectra of the second derivative for estimating the position of the maximum of the hidden component $E_d$ in the low-energy region.

Independent spectral data on the luminescence of InP/ZnS confirm the observed pattern. We made appropriate estimates for the halfwidth of the exciton band using data from the papers [23–27] (see Table 1 and Fig. 2). It can be seen that a noticeable increase in the halfwidth $H_x(T)$ with increasing temperature takes place only in [23]. All the other works demonstrate a temperature independent $H_x$. It is important to point out that, in [24–27], the value of $H_x$ at room temperature for 2 – 4.7 nm InP/ZnS nanocrystals varies within a wide range of 166 – 372 meV. The above effects are presumably due to different synthesis protocols that lead to the formation of a more or less wide size distribution of QDs, as well as the appearance of defects of a certain type in them.

**Table 1** – Spectral characteristics of exciton and defect-related luminescence in InP/ZnS nanocrystals

| Size, nm | $E_x$, eV | $H_x$, meV | $E_d$, eV | Temperature range, K | Precursors, Injection temperature, Manufacturer | Reference |
|---|---|---|---|---|---|---|
| 2.1 (QD-1) | 2.36 – 2.32 | 231 | 2.04 | 6.5 – 296 | $InX_3$ (X = Cl, Br, I), tris(diethylamino) phosphine, 180 °C, NIIPA, Russia | This work |
| 2.3 (QD-2) | 2.16 – 2.13 | 315 | 1.80 | | | |
| 1.8 | 2.45 – 2.35 | 165 – 236 | – | 2 – 300 | $In(Ac)_3$ tris(trimethylsilyl) phosphine, 188 °C | [23] |
| 3.0 | 2.14 – 2.05 | 134 – 217 | – | | | |
| – | 2.05 – 2.00 | 192 | 1.80 | 15 – 300 | $In(Ac)_3$, $PH_3$; 290 °C | [24] |
| 2.9 | 2.02 – 1.95 | 166 | 1.66 | 4 – 290 | $InCl_3$ tris(dimethylamino) phosphine, 160 °C | [25] |
| 2.9 | 2.08 – 2.06 | 229 | 1.78 | 11 – 300 | Mesolight Inc., China | [26] |
| 2 | 2.38 – 2.35 | 316 | 2.00 | 80 – 300 | Janus New-Materials Co. Ltd., China | [27] |
| 2.4 | 2.18 – 2.15 | 257 | 1.78 | | | |
| 2.9 | 1.84 – 1.83 | 310 | 1.66 | | | |
| 4.7 | 1.79 – 1.77 | 372 | – | | | |

Table 1 includes some information on the peculiarities of the fabrication of InP/ZnS nanocrystals. All of these techniques can be classified as a hot-injection one [38]. For each of the samples, the sources of indium and phosphorus and the injection temperature are given, respectively. So, the present work and [25] utilize relatively safe amine-derived sources of phosphorus for preparing QDs [31, 39]. To facilitate the subsequent growth of the shell and reduce the size distribution of the nanocrystals, $ZnCl_2$ is added to the initial mixture that can lead to the formation of Zn-doped InP core [31]. QDs in [23, 24] were synthesized using problematic, from the standpoint of green and economic chemistry, tris(trimethylsilyl)phosphine and $PH_3$ gas. In all the cases, InP core of QDs is characterized by a low luminescence quantum yield (< 1%). However, the latter significantly increases through coating with a shell. This fact evidences the relationship between non-radiative relaxation channels of excitations and the surface of the nanocrystals. Thus, it can be inferred that, whatever the synthesis peculiarities, inhomogeneous band broadening processes in the photoluminescence spectra are inherent to the majority of InP/ZnS quantum dots.

*4.3. Temperature shift of the exciton band*

Fig. 3 displays the temperature shift of the maximum of $E_x$ for the QDs explored. The experimental data were analyzed using Fan expression [35, 40–43]:

$$E_x(T) = E_x(0) - A_F n_s, \text{ where } n_s = \left[\exp\left(\frac{\hbar\omega}{kT}\right) - 1\right]^{-1}. \quad (1)$$

Here $E_x(0)$ is the energy of the PL maximum at 0 K; $A_F$ is the Fan parameter dependent on the microscopic properties of the material; $n_s$ is the Bose-Einstein factor for phonons with an average energy $\hbar\omega$; $k$ is the Boltzmann constant, eV/K. The approximations are shown in Fig. 3 by solid lines. The calculated values of the parameters are listed in Table 2.

**Table 2** – Parameters of exciton-phonon interaction in InP/ZnS

| Sample | $E_x(0)$, ± 0.01 eV | $A_F$, ± 3 meV | $\hbar\omega$, ±1 meV | $S$ | $E_{ss}$ at 6.5 K, ±10 meV |
|---|---|---|---|---|---|
| QD-1 | 2.35 | 38 | 11 | 1.73 | 359 |
| QD-2 | 2.16 | 12 | 9 | 0.66 | 332 |
| bulk InP [44] | 1.42 | 49 | 14 | 1.78 | – |

The effective phonon energies $\hbar\omega$ obtained in the experiment are in good agreement with the longitudinal acoustic vibrations of 10.2 meV at the L point of the Brillouin zone for bulk InP [45], as well as with our estimates based on the data about the temperature behavior of exciton luminescence in indium phosphide single crystals [44]. The ascertained fact satisfies the independent experimental results in [22, 23, 46] and the theoretical conclusions of [47–50] on the enhancement of the nonpolar interaction along the mechanism of the deformation potential between excitons and acoustic vibrations in the quantum confinement regime. It is known that low-frequency modes with an energy of ≈ 18.6 meV are identified in the Raman spectra of InP nanocrystals modified by treating in HF, doping with Zn, or coating with a ZnS shell [33]. Thus, the temperature shift of the exciton luminescence band in the InP/ZnS samples investigated is caused by the exciton-phonon interaction with longitudinal modes of acoustic vibrations.

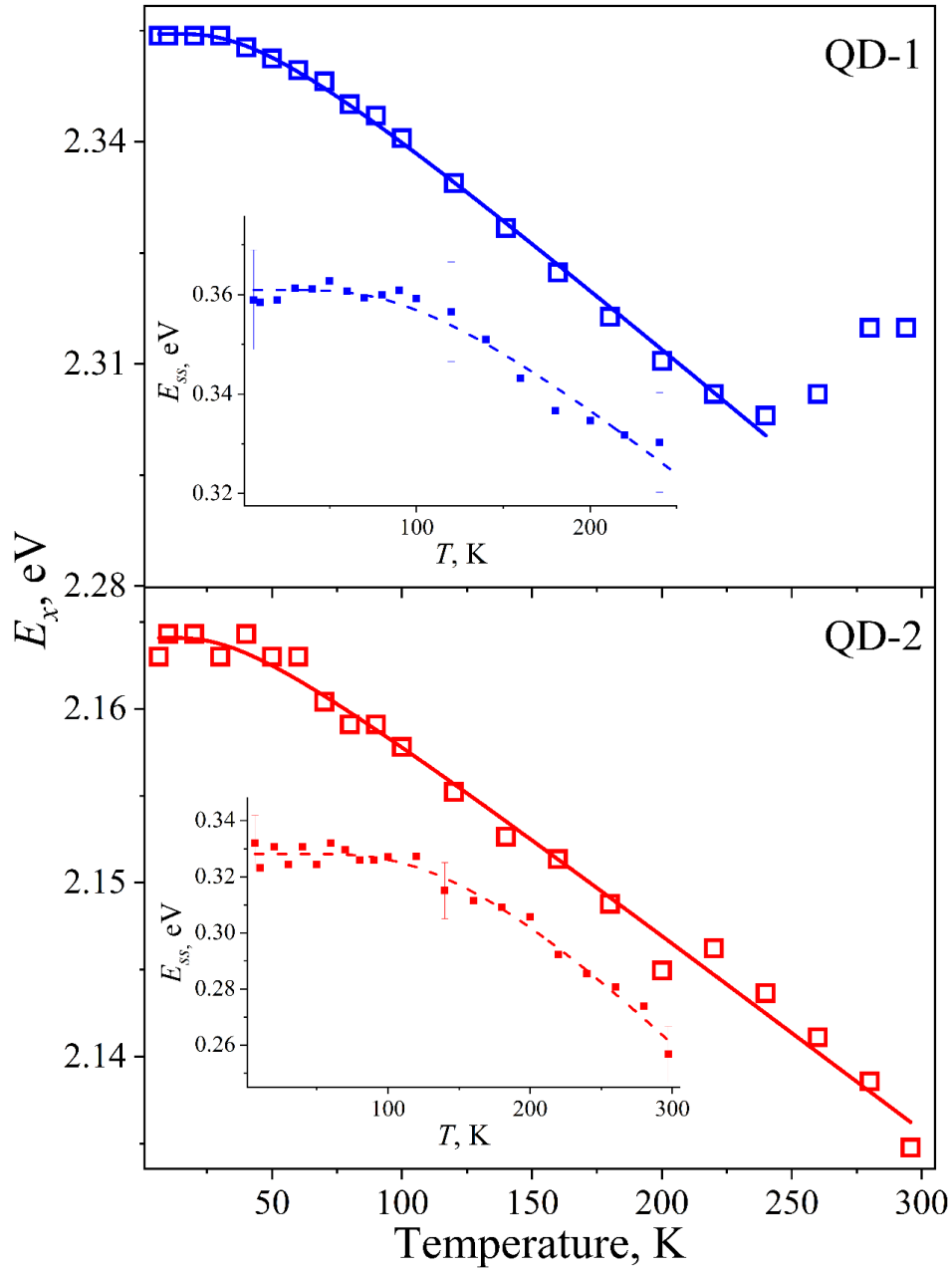

**Figure 3.** Temperature dependencies for the maximum exciton emission $E_x$ (open squares) and Stokes shift $E_{ss}$ (insert, solid squares) of the InP/ZnS samples under study. Symbols are experimental data, solid lines designate approximations according to Fan expression (1), dashed lines indicate free-hand curves.

Relying on the values for the maxima of the exciton band in the absorption $E_1$ [30] and emission $E_x$ (the present work) spectra we can analyze the temperature dependencies of the Stokes shift $E_{ss} = E_1 - E_x$ in the QDs (see inserts in Fig. 3). It can be seen that $E_{ss}$ remains intact at low temperatures but slopes down with a further increase in temperature. The $E_{ss}(T)$ dependence appears to be revealed due to a fine structure of excited exciton levels, typical for QDs based on various semiconductor compounds, including CdSe [47, 51, 52], CdTe/CdS [53], InP [54], InP/ZnS [25, 52]. The quantum confinement effect enhances the exchange interaction. As a consequence, the splitting energy $\Delta E$ between the bright *2* and dark *1* exciton states becomes higher. At low temperatures, the dark state dominates in the population distribution over exciton fine structure levels. Heating leads to a redistribution of excited carriers in favor of the high-energy bright state. Consequently, the emission energy $E_x$ increases, and the $E_{ss}$ diminishes. The Stokes shift is also affected by the exciton-phonon interaction that can be estimated using the Huang-Rhys factor $S = A_F/2\hbar\omega$ [28]. It should be stressed that the value of the Stokes shift is larger for QD-1, which is consistent with the obtained estimates of $S$ (see Table 2). An increase in the value of $E_{ss}$ with decreasing nanocrystal size is a usual regularity previously found for QDs based on InP [22, 23], as well as for CdTe [55], CdSe [47], and CdSe/ZnS [56]. Note that the experiments performed do not allow to distinguish the contributions of exciton fine structure and exciton-phonon coupling to the Stokes shift.

*4.4 Thermal quenching of exciton and defect-related emission*

The efficiency of radiative recombination processes for the QDs tested is strongly temperature-dependent. Fig. 4 sketches the temperature dependencies for the integrated intensities $I_x$ and $I_d$ after normalization to their maximum values. The analyzed spectral ranges associated with excitonic and defect-related luminescence are appropriately colored in the Fig. 1.

For accounting for the relationship between the probabilities of radiative and non-radiative transitions between discrete energy levels, the conventional model of thermal quenching of photoluminescence in solids can be presented by the well-known Mott expression [57–59]:

$$I(T) = I_0 \eta(T, E_q), \quad \eta(T, E_q) = \left[1 + p \exp\left(-\frac{E_q}{kT}\right)\right]^{-1}, \qquad (2)$$

where the function $\eta(T, E_q)$ characterizes the efficiency of radiative transitions; $I_0$ is the luminescence intensity without quenching, a.u.; $p$ is a dimensionless pre-exponential factor; $E_q$ is the activation energy of quenching, eV. In the case of involving several channels of non-radiative relaxation of excitation in the quenching mechanism, the efficiency function acquires additional temperature summands with the $p$ and $E_q$ parameters [60–63]. Fig. 4 (see the insert) displays the experimental data for QD-1 in Arrhenius coordinates. Colored solid lines indicate the results of simulation using the expression (2) under the assumption of a different number $N$ of non-radiative channels. It can be seen that an increase in this quantity improves the description accuracy, and the best one is achieved for $N = 10$. In doing so, it is obvious that the model of discrete energy levels with such a large number of adjustable parameters prevents from interpreting a multiplicity of thermally activated barriers that can form the temperature dependencies observed.

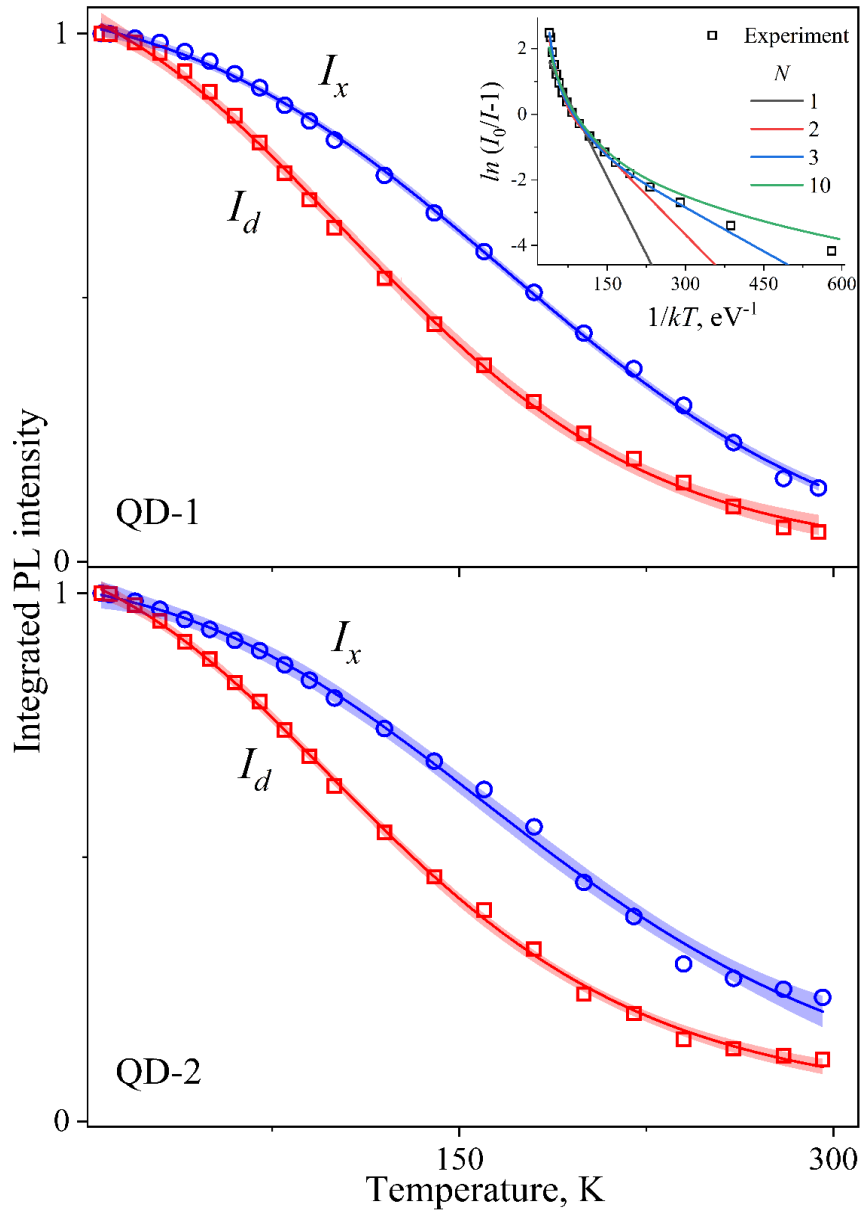

**Figure 4.** PL temperature dependencies for the InP/ZnS samples.
Symbols denote experimental data, solid lines are calculated curves according to the expression (2) accounting for the distribution of the activation energy, blue and red shaded areas along the lines indicate 99% confidence intervals. Insert shows measured (open squares) and simulation (colored lines) data in Arrhenius coordinates, see section 4.4 for details.

Let us look at a band model for describing the possible thermal quenching mechanisms for luminescence of various origin in excited InP/ZnS nanocrystals. A schematic representation of the energy levels involved in the emission processes is presented in Fig. 5a. The alignment of the InP and ZnS energy bands leads to the formation of type-I core/shell quantum dots. Synthesis may facilitate the emergence of $Zn_{In}$ substitutional defects in the InP core [33] and dangling bonds of indium $DB_{In}$ and phosphorus $DB_P$ at the core/shell interface due to incomplete passivation. In turn, these bonds create the corresponding donor and acceptor levels [20]. In particular, dangling bonds of phosphorus atoms on the surface of InP nanocrystals have been previously uncovered by high-resolution photoelectron spectroscopy using synchrotron radiation [64].

The recombination of excited charge carriers from the *2* and *1* levels to the *0* ground state forms the $E_x$ emission band (see Fig. 5b). Non-radiative channels associated with surface states produce a low quantum yield in shell-free nanocrystals [65–67]. With thermal activation, the quenching of exciton luminescence in the quantum dots is possible due to the escape of an electron from the InP core to the ZnS shell. Similar processes are typical for epitaxial QDs when carriers thermally escaped into different layers of the heterostructure are captured by non-radiative recombination centers [15, 68, 69].

The described level structure is usual for an individual nanocrystal. Simultaneously, the samples explored are ensembles of QDs of various sizes. Hence, an energy distribution of exciton levels takes place: with a decline in the core size, both the energy between the ground and dark states and the gap $\Delta E$ in the doublet of the dark and bright states increase. In turn, the energy position of the bottom of the conduction band (CB) in the shell is governed by the latter's thickness. The height of the $E_q$ activation barrier of a non-radiative channel for excited electrons to pass into the shell is assumed to be described by a distribution density, $f(E_q)$. Thus, to analyze the exciton PL quenching processes in the QDs, we resort to an equivalent band scheme (Fig. 5b, right), making an allowance for the distribution $f(E_q)$.

The defect-related emission $E_d$ arises as a result of the transitions shown in the diagram in the left panel of Fig. 5c. In this case, the PL quenching may be as a consequence of a thermally activated transition of electrons from the defect $DB_{In}$-state to the *2* and *1* levels or, respectively, holes from $Zn_{In}$ and $DB_P$ to the *0* ground level, followed by participation in the exciton emission mechanisms. Note that the different quality of the core/shell interface and location of the impurity zinc atom in the lattice is the reason for the energy levels of the $DB_{In}$, $DB_P$, and $Zn_{In}$ defects to be distributed in a certain range of values [20, 33]. Then, the equivalent band diagram proposed in the right panel of Fig. 5c helps run an analysis of the experimentally observed quenching of defect-related luminescence. Here, the activation energy of quenching is related to the effective-acceptor-level depth and can also be described by the $f(E_q)$ distribution density.

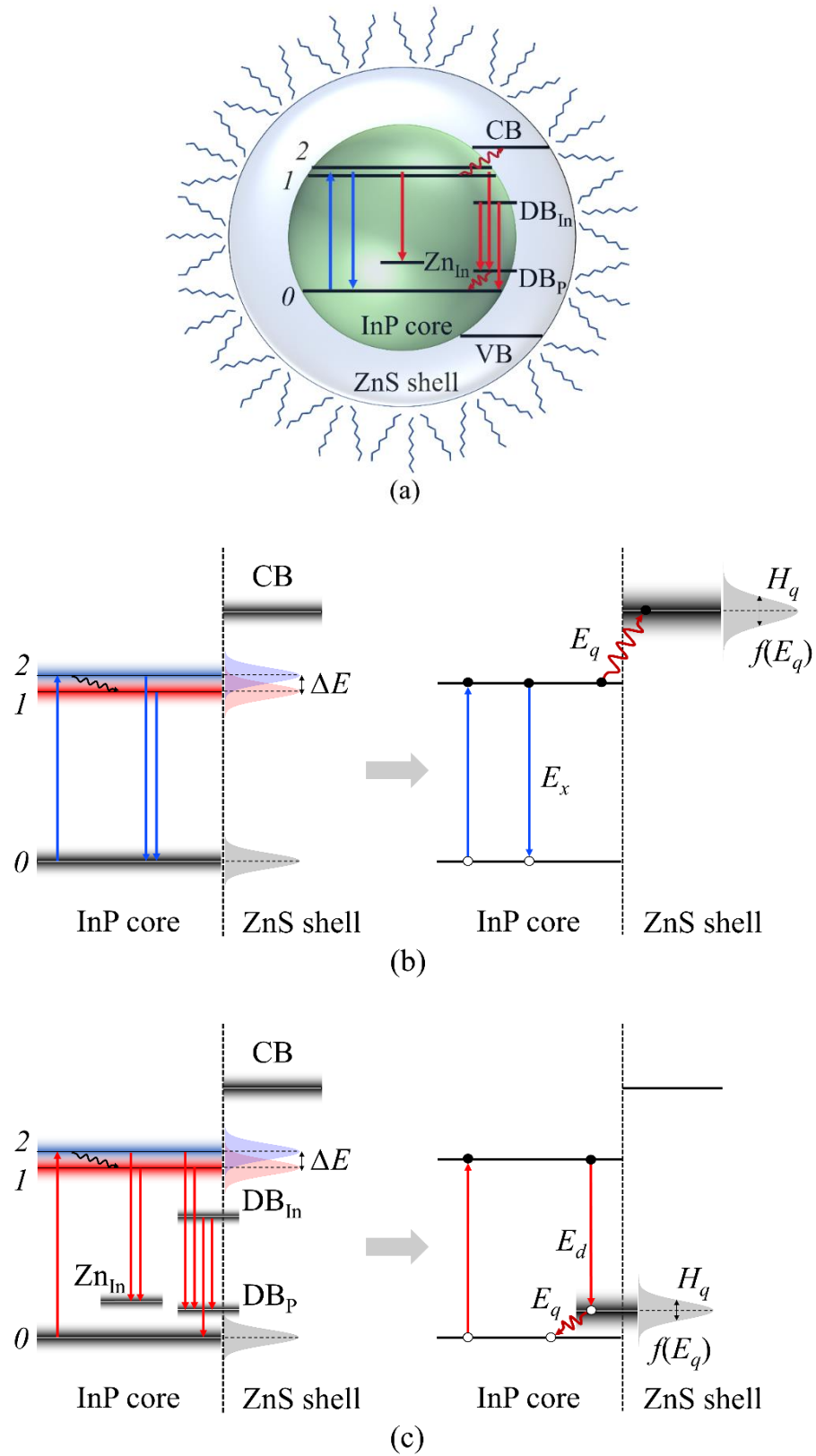

**Figure 5.** Band diagrams for the InP/ZnS nanocrystals (a) describing the mechanisms of quenching of the exciton (b) and defect-related (c) luminescence. See section 4.4 for the discussion of the energy levels.

Exploiting the above equivalent band schemes and accounting for the efficiency of radiative transition function $\eta(T, E_q)$ and the $f(E_q)$ distribution of the activation energy of quenching [70–73], we can write down the temperature dependence of the PL intensity:

$$I(T) = I_0 \int_0^E \eta(T, E_q) f(E_q) dE_q,$$

$$f(E_q) = (4\ln 2)^{\frac{1}{2}} \pi^{-\frac{1}{2}} H_q^{-1} e^{\left(-4\ln 2 (E_q - E_{qm})^2 H_q^{-2}\right)}.$$

(3)

Approximating the experimental data, we suppose that the distribution function has a Gaussian shape with a maximum energy $E_{qm}$ and a halfwidth $H_q$. Bear in mind that the model does not consider the temperature dependence of the $f(E_q)$ parameters.

*4.5. Activation energy distributions*

The approximation of the experimental data for the thermal quenching of exciton and defect-related luminescence along the expression (3) is shown in Fig. 4 (solid lines). Appropriate colors mark 99%-confidence intervals in the derived curves. It can be seen that the approach proposed well approximates the experimental dependencies (Adj. R-Square > 0.997). The $f(E_q)$ functions deduced from the calculations for the QDs are outlined in Fig. 6. Their parameters are given in Table 3. It is worth stressing that the defect-related PL for both samples is characterized by lower magnitudes of $E_{qm}$ and $H_q$. The predicted model distributions $f(E_q)$ have a nonzero density of states for $E_q = 0$. From the point of view of physical interpretation, it can be claimed that non-activation (barrier-free) quenching processes proceed in the nanocrystals, presumably due to tunneling [74, 75]. In this case, the barrier-free processes are more typical for thermal quenching of the defect-related emission.

The barrier height of $\Delta E_{\text{InP/ZnS}} = 500$ meV in the InP/ZnS bulk heterojunction meets the difference in electron affinity energies of 4.4 eV for indium phosphide and 3.9 eV for zinc sulfide [23, 76]. In the case of a nanostructure, the size effect affects the position of the energy levels [77]. According to our estimates within the approach of [20, 78] for the nanocrystals studied, the $\Delta E_{\text{InP/ZnS}}$ barrier for the escape of an electron into the shell decreases to 100 – 320 meV. The above boundary values are in good agreement with the ranges calculated for the distribution of the activation energy for quenching of exciton luminescence in QD-1 and QD-2 (see Fig. 6, blue dashed lines). In turn, the maximum of the obtained distribution of the activation energy of quenching for defect-related emission is quite well matches the independent theoretical estimate of 86 meV for the distance between the *0* and DB$_P$ energy levels in a partially passivated cluster

(InP)$_{34}$(ZnS)$_{46}$ [20] (see the red dashed line in Fig. 6). Noteworthy is that $H_q > H_x$ for exciton emission in both samples. This result can be qualitatively interpreted in the framework of a quantum mechanical particle-in-a-box model [1, 3] for confined electron-hole excitations. In this case, the radiative relaxation of the exciton is mainly determined by the properties (size, shape, etc.) of the core, which form the energy parameters of the corresponding quantum well. On the other hand, the process of exciton emission quenching during the electron thermal escape over the $\Delta E_{\text{InP/ZnS}}$ activation barrier will be governed by the spread of the parameters of the core and the shell alike. Therefore, one should expect a wider distribution for $f(E_q)$, and it is exactly what we are observing within the estimates made.

**Table 3** – Model parameters of thermal quenching for exciton and defect-related luminescence in the InP/ZnS

| Sample | Emission | $E_{qm}$, ± 10 meV | $H_q$, ±20 meV | $p$ | Quality parameter δ, ± 1.0 % |
|---|---|---|---|---|---|
| QD-1 | exciton | 220 | 310 | 2.49·10$^6$ | 12.1 quenching<br>9.8 emission<br>10.7 absorption [30]<br>11.1 size [30] |
| | defect-related | 69 | 120 | 1.13·10$^3$ | – |
| QD-2 | exciton | 224 | 376 | 6.22·10$^6$ | 15.6 quenching<br>14.6 emission<br>15.1 absorption [30]<br>17.4 size [30] |
| | defect-related | 52 | 90 | 1.61·10$^2$ | – |

Earlier [30], we have demonstrated that the quality of the size distribution of QDs in an ensemble can be quantitatively analyzed using the parameter δ. The latter is estimated as the ratio of the halfwidth to the distribution maximum and properly correlates with the relative halfwidth δ = $H/E_m$ of the optical absorption and luminescence bands. Note that, when calculated by the spectral bands of a Gaussian shape, the parameter δ is directly related to the coefficient of variation for the energy distribution of the appropriate optical transitions [79]. The current work computes values of δ for the exciton PL bands and for the $f(E_q)$ distributions of the quenching activation energy relative to the *0* state. The findings secured are in good agreement with each other within

QD ensemble, see Table 3. This fact indicates a strong correlation between the considered spectral and size distributions. Noteworthy, the δ values for defect-related photoluminescence do not exhibit such a correlation.

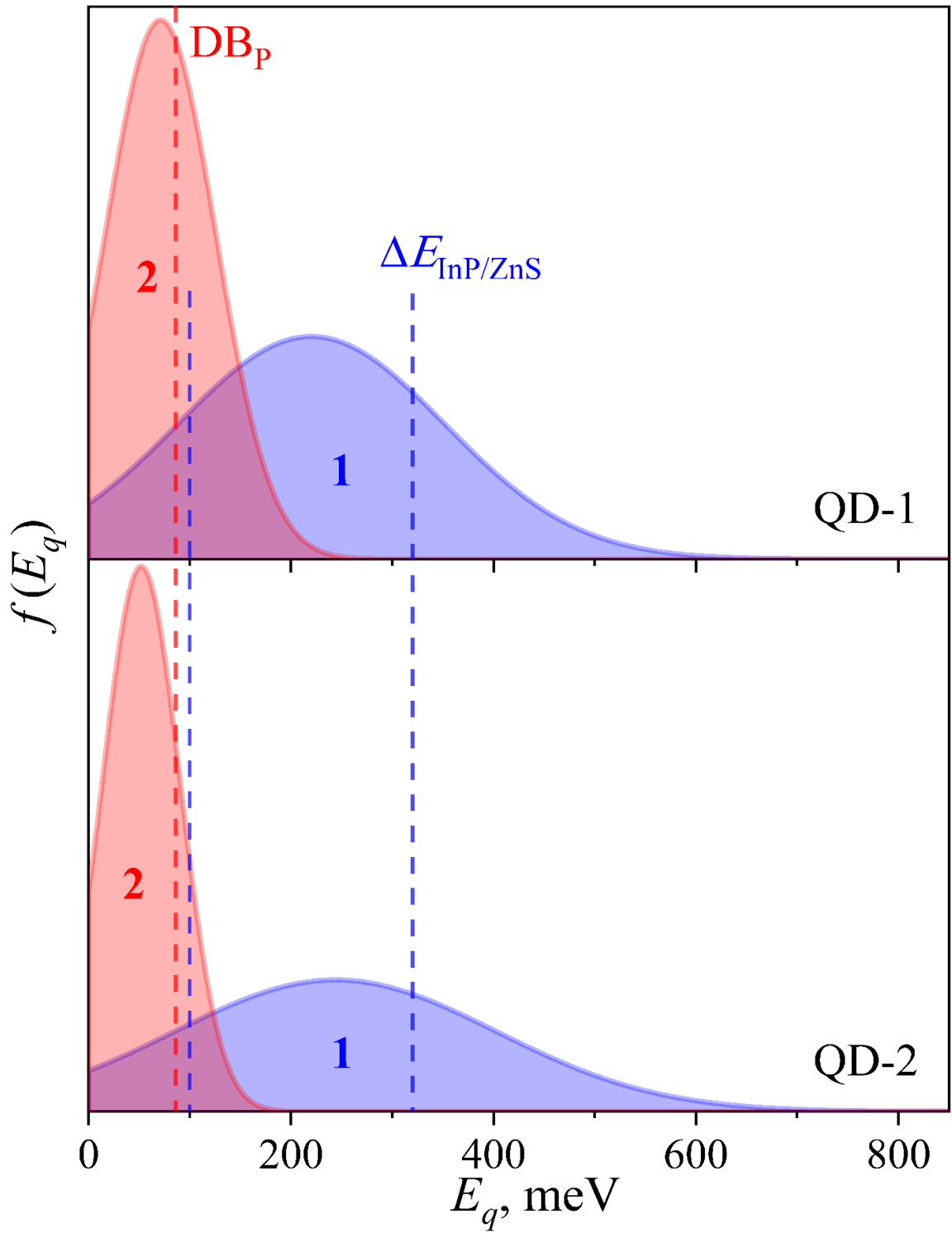

**Figure 6.** Model distribution functions of the activation energy for the quenching mechanisms of exciton (1) and defect-related (2) PL in the InP/ZnS nanocrystals. Dashed lines label theoretical estimates of energy barriers related to InP/ZnS heterojunction and to levels for dangling bonds of phosphorus, see section 4.5 for details.

## 5. Conclusion

We have studied the mechanisms of thermal quenching of PL in ensembles of water-soluble environmentally friendly InP/ZnS quantum dots with an average particle size of 2.1 (QD-1) and 2.3 (QD-2) nm. Examining the spectra evolved in the range from 296 to 6.5 K, we can infer that the emission composition forms by two bands, $E_x$ and $E_d$. The latter are due to transitions involving exciton and defect states, respectively. In the process, the emission intensity increases by a factor of 7 and 18, respectively, as the temperature goes down from 296 to 6.5 K. The observable temperature shift of the $E_x$ band is a consequence of exciton-phonon interaction with longitudinal modes of acoustic vibrations. Comparison with the optical absorption data identifies a temperature dependence of the Stokes shift. This fact confirms the existence of a fine structure of the exciton emission. The inhomogeneous broadening of the $E_x$ exciton luminescence band is a result of the spread of nanocrystals in size, shape, and other structural and physicochemical parameters. The conducted analysis of the PL quenching processes evidences a distribution of the activation energy of non-radiative relaxation of excitations for each luminescence band. In the paper, we have proposed band schemes and mechanisms for quenching of exciton and defect-related emission using the $f(E_q)$ Gaussian distribution of the activation energy. The parameters calculated for the model distributions (maximum $E_{qm}$; halfwidth $H_q$) amount to (220 meV; 310 meV) in QD-1 and (224 meV; 376 meV) in QD-2 for the $E_x$ band and (69 meV; 120 meV) in QD-1 and (52 meV; 90 meV) in QD-2 for the $E_d$ band. The findings secured indicate that the thermal escape of an electron from the InP core to the ZnS shell is dominant in the mechanisms of non-radiative decay of excitons. The quenching of the defect-related luminescence can be quantitatively described within an equivalent band model to take into account the processes of thermally activated transitions of holes from the level corresponding to dangling phosphorus $DB_P$ bonds. We have demonstrated that these transitions are realized not only due to thermal activation but also by tunneling. In addition, analyzing the quality parameter δ we have revealed a correlation between size distributions in QDs ensemble, function $f(E_q)$ and relative spectral halfwidth of exciton absorption and emission bands. The research held broadens the insight into the mechanisms of thermal quenching of photoluminescence in InP-based type-I nanocrystals and can be used for optimizing the methods of their directed synthesis, aimed at enhancing the efficiency of radiative processes.

**Acknowledgements**

The work was supported by Minobrnauki research project FEUZ-2020-0059.

# References

[1] H. Weller, Colloidal Semiconductor Q-Particles - Chemistry in the Transition Region Between Solid-State and Molecules, Angew. Chemie-International Ed. English. 32 (1993) 41–53. https://doi.org/10.1002/anie.199300411.

[2] A.P. Alivisatos, Semiconductor Clusters, Nanocrystals, and Quantum Dots, Science. 271 (1996) 933–937. https://doi.org/10.1126/science.271.5251.933.

[3] A.D. Yoffe, Low-dimensional systems: quantum size effects and electronic properties of semiconductor microcrystallites (zero-dimensional systems) and some quasi-two-dimensional systems, Adv. Phys. 42 (1993) 173–262. https://doi.org/10.1080/00018739300101484.

[4] X. Michalet, Quantum Dots for Live Cells, in Vivo Imaging, and Diagnostics, Science. 307 (2005) 538–544. https://doi.org/10.1126/science.1104274.

[5] P. V. Kamat, Quantum Dot Solar Cells. Semiconductor Nanocrystals as Light Harvesters, J. Phys. Chem. C. 112 (2008) 18737–18753. https://doi.org/10.1021/jp806791s.

[6] X. Dai, Z. Zhang, Y. Jin, Y. Niu, H. Cao, X. Liang, L. Chen, J. Wang, X. Peng, Solution-processed, high-performance light-emitting diodes based on quantum dots, Nature. 515 (2014) 96–99. https://doi.org/10.1038/nature13829.

[7] V.I. Klimov, Optical Gain and Stimulated Emission in Nanocrystal Quantum Dots, Science. 290 (2000) 314–317. https://doi.org/10.1126/science.290.5490.314.

[8] J.M. Pietryga, Y.-S. Park, J. Lim, A.F. Fidler, W.K. Bae, S. Brovelli, V.I. Klimov, Spectroscopic and Device Aspects of Nanocrystal Quantum Dots, Chem. Rev. 116 (2016) 10513–10622. https://doi.org/10.1021/acs.chemrev.6b00169.

[9] G. Konstantatos, I. Howard, A. Fischer, S. Hoogland, J. Clifford, E. Klem, L. Levina, E.H. Sargent, Ultrasensitive solution-cast quantum dot photodetectors, Nature. 442 (2006) 180–183. https://doi.org/10.1038/nature04855.

[10] S.S. Savchenko, A.S. Vokhmintsev, I.A. Weinstein, Luminescence parameters of InP/ZnS@AAO nanostructures, AIP Conf. Proc. 1717 (2016) 040028. https://doi.org/10.1063/1.4943471.

[11] S.S. Savchenko, A.S. Vokhmintsev, I.A. Weinstein, Optical properties of InP/ZnS quantum dots deposited into nanoporous anodic alumina, J. Phys. Conf. Ser. 741 (2016) 012151. https://doi.org/10.1088/1742-6596/741/1/012151.

[12] A. Bednarkiewicz, J. Drabik, K. Trejgis, D. Jaque, E. Ximendes, L. Marciniak, Luminescence based temperature bio-imaging: Status, challenges, and perspectives, Appl. Phys. Rev. 8 (2021) 011317. https://doi.org/10.1063/5.0030295.

[13] M. Gonschorek, H. Schmidt, J. Bauer, G. Benndorf, G. Wagner, G.E. Cirlin, M. Grundmann, Thermally assisted tunneling processes in InxGa1-xAs/GaAs quantum-dot structures, Phys. Rev. B. 74 (2006) 115312. https://doi.org/10.1103/PhysRevB.74.115312.

[14] Y. Zhao, C. Riemersma, F. Pietra, R. Koole, C. de Mello Donegá, A. Meijerink, High-Temperature Luminescence Quenching of Colloidal Quantum Dots, ACS Nano. 6 (2012) 9058–9067. https://doi.org/10.1021/nn303217q.

[15]    T. V. Torchynska, J.L. Casas Espinola, L. V. Borkovska, S. Ostapenko, M. Dybiec, O. Polupan, N.O. Korsunska, A. Stintz, P.G. Eliseev, K.J. Malloy, Thermal activation of excitons in asymmetric InAs dots-in-a-well InxGa1−xAs∕GaAs structures, J. Appl. Phys. 101 (2007) 024323. https://doi.org/10.1063/1.2427105.

[16]    D. Valerini, A. Cretí, M. Lomascolo, L. Manna, R. Cingolani, M. Anni, Temperature dependence of the photoluminescence properties of colloidal CdSe/ZnS core/shell quantum dots embedded in a polystyrene matrix, Phys. Rev. B. 71 (2005) 235409. https://doi.org/10.1103/PhysRevB.71.235409.

[17]    S.A. Crooker, T. Barrick, J.A. Hollingsworth, V.I. Klimov, Multiple temperature regimes of radiative decay in CdSe nanocrystal quantum dots: Intrinsic limits to the dark-exciton lifetime, Appl. Phys. Lett. 82 (2003) 2793–2795. https://doi.org/10.1063/1.1570923.

[18]    P. Jing, J. Zheng, M. Ikezawa, X. Liu, S. Lv, X. Kong, J. Zhao, Y. Masumoto, Temperature-Dependent Photoluminescence of CdSe-Core CdS/CdZnS/ZnS-Multishell Quantum Dots, J. Phys. Chem. C. 113 (2009) 13545–13550. https://doi.org/10.1021/jp902080p.

[19]    C.E. Rowland, W. Liu, D.C. Hannah, M.K.Y. Chan, D. V. Talapin, R.D. Schaller, Thermal stability of colloidal InP nanocrystals: Small inorganic ligands boost high-temperature photoluminescence, ACS Nano. 8 (2014) 977–985. https://doi.org/10.1021/nn405811p.

[20]    E. Cho, T. Kim, S. Choi, H. Jang, K. Min, E. Jang, Optical Characteristics of the Surface Defects in InP Colloidal Quantum Dots for Highly Efficient Light-Emitting Applications, ACS Appl. Nano Mater. 1 (2018) 7106–7114. https://doi.org/10.1021/acsanm.8b01947.

[21]    K.E. Hughes, J.L. Stein, M.R. Friedfeld, B.M. Cossairt, D.R. Gamelin, Effects of Surface Chemistry on the Photophysics of Colloidal InP Nanocrystals, ACS Nano. 13 (2019) 14198–14207. https://doi.org/10.1021/acsnano.9b07027.

[22]    A. Narayanaswamy, L.F. Feiner, P.J. van der Zaag, Temperature Dependence of the Photoluminescence of InP/ZnS Quantum Dots, J. Phys. Chem. C. 112 (2008) 6775–6780. https://doi.org/10.1021/jp800339m.

[23]    A. Narayanaswamy, L.F. Feiner, A. Meijerink, P.J. van der Zaag, The Effect of Temperature and Dot Size on the Spectral Properties of Colloidal InP/ZnS Core−Shell Quantum Dots, ACS Nano. 3 (2009) 2539–2546. https://doi.org/10.1021/nn9004507.

[24]    T.T. Pham, T.K. Chi Tran, Q.L. Nguyen, Temperature-dependent photoluminescence study of InP/ZnS quantum dots, Adv. Nat. Sci. Nanosci. Nanotechnol. 2 (2011) 025001. https://doi.org/10.1088/2043-6262/2/2/025001.

[25]    L. Biadala, B. Siebers, Y. Beyazit, M.D. Tessier, D. Dupont, Z. Hens, D.R. Yakovlev, M. Bayer, Band-Edge Exciton Fine Structure and Recombination Dynamics in InP/ZnS Colloidal Nanocrystals, ACS Nano. 10 (2016) 3356–3364. https://doi.org/10.1021/acsnano.5b07065.

[26]    T. Chen, K. Li, H. Mao, Y. Chen, J. Wang, G. Weng, Photoluminescence Investigation of the InP/ZnS Quantum Dots and Their Coupling with the Au Nanorods, J. Electron. Mater. 48 (2019) 3497–3503. https://doi.org/10.1007/s11664-019-07106-9.

[27]    C. Wang, Q. Wang, Z. Zhou, W. Wu, Z. Chai, Y. Gao, D. Kong, Temperature dependence of photoluminescence properties in InP/ZnS core-shell quantum dots, J. Lumin. 225 (2020) 117354. https://doi.org/10.1016/j.jlumin.2020.117354.

[28]    S.S. Savchenko, A.S. Vokhmintsev, I.A. Weinstein, Temperature-induced shift of the exciton absorption band in InP/ZnS quantum dots, Opt. Mater. Express. 7 (2017) 354. https://doi.org/10.1364/OME.7.000354.

[29]    S. Savchenko, A. Vokhmintsev, I. Weinstein, Exciton–Phonon Interactions and Temperature Behavior of Optical Spectra in Core/Shell InP/ZnS Quantum Dots, in: X. Tong, Z.M. Wang (Eds.), Core/Shell Quantum Dots, Springer, 2020: pp. 165–196. https://doi.org/10.1007/978-3-030-46596-4_5.

[30]    S.S. Savchenko, I.A. Weinstein, Inhomogeneous Broadening of the Exciton Band in Optical Absorption Spectra of InP/ZnS Nanocrystals, Nanomaterials. 9 (2019) 716. https://doi.org/10.3390/nano9050716.

[31]    M.D. Tessier, D. Dupont, K. De Nolf, J. De Roo, Z. Hens, Economic and Size-Tunable Synthesis of InP/ZnE (E = S, Se) Colloidal Quantum Dots, Chem. Mater. 27 (2015) 4893–4898. https://doi.org/10.1021/acs.chemmater.5b02138.

[32]    M. Grabolle, M. Spieles, V. Lesnyak, N. Gaponik, A. Eychmüller, U. Resch-Genger, Determination of the Fluorescence Quantum Yield of Quantum Dots: Suitable Procedures and Achievable Uncertainties, Anal. Chem. 81 (2009) 6285–6294. https://doi.org/10.1021/ac900308v.

[33]    E.M. Janke, N.E. Williams, C. She, D. Zherebetskyy, M.H. Hudson, L. Wang, D.J. Gosztola, R.D. Schaller, B. Lee, C. Sun, G.S. Engel, D. V. Talapin, Origin of Broad Emission Spectra in InP Quantum Dots: Contributions from Structural and Electronic Disorder, J. Am. Chem. Soc. 140 (2018) 15791–15803. https://doi.org/10.1021/jacs.8b08753.

[34]    S.S. Savchenko, A.S. Vokhmintsev, I.A. Weinstein, Effect of temperature on the spectral properties of InP/ZnS nanocrystals, J. Phys. Conf. Ser. 961 (2018) 012003. https://doi.org/10.1088/1742-6596/961/1/012003.

[35]    S.S. Savchenko, A.S. Vokhmintsev, I.A. Weinstein, Photoluminescence thermal quenching of yellow-emitting InP/ZnS quantum dots, AIP Conf. Proc. 2015 (2018) 020085. https://doi.org/10.1063/1.5055158.

[36]    S.B. Brichkin, M.G. Spirin, S.A. Tovstun, V.Y. Gak, E.G. Mart'yanova, V.F. Razumov, Colloidal quantum dots InP@ZnS: Inhomogeneous broadening and distribution of luminescence lifetimes, High Energy Chem. 50 (2016) 395–399. https://doi.org/10.1134/S0018143916050064.

[37]    S.S. Savchenko, A.S. Vokhmintsev, I.A. Weinstein, Spectral Features and Luminescence Thermal Quenching of InP/ZnS Quantum Dots within 7.5 – 295 K Range, in: Adv. Photonics 2018 (BGPP, IPR, NP, NOMA, Sensors, Networks, SPPCom, SOF), OSA, Washington, D.C., 2018: p. NoW1J.4. https://doi.org/10.1364/NOMA.2018.NoW1J.4.

[38]    B. Chen, D. Li, F. Wang, InP Quantum Dots: Synthesis and Lighting Applications, Small. 16 (2020) 2002454. https://doi.org/10.1002/smll.202002454.

[39]    W.-S. Song, H.-S. Lee, J.C. Lee, D.S. Jang, Y. Choi, M. Choi, H. Yang, Amine-derived synthetic approach to color-tunable InP/ZnS quantum dots with high fluorescent qualities, J. Nanoparticle Res. 15 (2013) 1750. https://doi.org/10.1007/s11051-013-1750-y.

[40]    H.Y. Fan, Temperature dependence of the energy gap in semiconductors, Phys. Rev. 82 (1951) 900–905. https://doi.org/10.1103/PhysRev.82.900.

[41]    I. Weinstein, A. Zatsepin, Y. Shchapova, The phonon-assisted shift of the energy levels of localized electron states in statically disordered solids, Phys. B Condens. Matter. 263–264 (1999) 167–169. https://doi.org/10.1016/S0921-4526(98)01213-7.

[42]    I.A. Vainshtein, A.F. Zatsepin, V.S. Kortov, Applicability of the empirical Varshni relation for the temperature dependence of the width of the band gap, Phys. Solid State. 41 (1999) 905–908. https://doi.org/10.1134/1.1130901.

[43]    K.R. Karimullin, A.I. Arzhanov, I.Y. Eremchev, B.A. Kulnitskiy, N. V. Surovtsev, A. V. Naumov, Combined photon-echo, luminescence and Raman spectroscopies of layered ensembles of colloidal quantum dots, Laser Phys. 29 (2019). https://doi.org/10.1088/1555-6611/ab4bdb.

[44]    W.J. Turner, W.E. Reese, G.D. Pettit, Exciton Absorption and Emission in InP, Phys. Rev. 136 (1964) 1955–1958. https://doi.org/10.1103/PhysRev.136.A1467.

[45]    G.F. Alfrey, P.H. Borcherds, Phonon frequencies from the Raman spectrum of indium phosphide, J. Phys. C Solid State Phys. 5 (1972) L275–L278. https://doi.org/10.1088/0022-3719/5/20/002.

[46]    U. Banin, G. Cerullo, A.A. Guzelian, C.J. Bardeen, A.P. Alivisatos, C. V. Shank, Quantum confinement and ultrafast dephasing dynamics in InP nanocrystals, Phys. Rev. B. 55 (1997) 7059–7067. https://doi.org/10.1103/PhysRevB.55.7059.

[47]    M. Nirmal, D.J. Norris, M. Kuno, M.G. Bawendi, A.L. Efros, M. Rosen, Observation of the "dark exciton" in CdSe quantum dots, Phys. Rev. Lett. 75 (1995) 3728–3731. https://doi.org/10.1103/PhysRevLett.75.3728.

[48]    S. Schmitt-Rink, D.A.B. Miller, D.S. Chemla, Theory of the linear and nonlinear optical properties of semiconductor microcrystallites, Phys. Rev. B. 35 (1987) 8113–8125. https://doi.org/10.1103/PhysRevB.35.8113.

[49]    D.M. Mittleman, R.W. Schoenlein, J.J. Shiang, V.L. Colvin, A.P. Alivisatos, C. V. Shank, Quantum size dependence of femtosecond electronic dephasing and vibrational dynamics in CdSe nanocrystals, Phys. Rev. B. 49 (1994) 14435–14447. https://doi.org/10.1103/PhysRevB.49.14435.

[50]    T. Takagahara, Electron-phonon interactions in semiconductor nanocrystals, J. Lumin. 70 (1996) 129–143. https://doi.org/10.1016/0022-2313(96)00050-6.

[51]    A.L. Efros, M. Rosen, M. Kuno, M. Nirmal, D.J. Norris, M. Bawendi, Band-edge exciton in quantum dots of semiconductors with a degenerate valence band: Dark and bright exciton states, Phys. Rev. B. 54 (1996) 4843–4856. https://doi.org/10.1103/PhysRevB.54.4843.

[52]    A. Franceschetti, H. Fu, L.W. Wang, A. Zunger, Many-body pseudopotential theory of excitons in InP and CdSe quantum dots, Phys. Rev. B. 60 (1999) 1819–1829. https://doi.org/10.1103/PhysRevB.60.1819.

[53]    T. Watanabe, K. Takahashi, K. Shimura, D. Kim, Influence of carrier localization at the core/shell interface on the temperature dependence of the Stokes shift and the photoluminescence decay time in CdTe/CdS type-II quantum dots, Phys. Rev. B. 96 (2017) 035305. https://doi.org/10.1103/PhysRevB.96.035305.

[54]    O.I. Mićić, H.M. Cheong, H. Fu, A. Zunger, J.R. Sprague, A. Mascarenhas, A.J. Nozik, Size-dependent spectroscopy of InP quantum dots, J. Phys. Chem. B. 101 (1997) 4904–4912. https://doi.org/10.1021/jp9704731.

[55]   J. Pérez-Conde, A.K. Bhattacharjee, M. Chamarro, P. Lavallard, V.D. Petrikov, A.A. Lipovskii, Photoluminescence Stokes shift and exciton fine structure in CdTe nanocrystals, Phys. Rev. B. 64 (2001) 113303. https://doi.org/10.1103/PhysRevB.64.113303.

[56]   A. Joshi, K.Y. Narsingi, M.O. Manasreh, E.A. Davis, B.D. Weaver, Temperature dependence of the band gap of colloidal CdSe∕ZnS core/shell nanocrystals embedded into an ultraviolet curable resin, Appl. Phys. Lett. 89 (2006) 131907. https://doi.org/10.1063/1.2357856.

[57]   M. Leroux, N. Grandjean, B. Beaumont, G. Nataf, F. Semond, J. Massies, P. Gibart, Temperature quenching of photoluminescence intensities in undoped and doped GaN, J. Appl. Phys. 86 (1999) 3721–3728. https://doi.org/10.1063/1.371242.

[58]   M.A. Reshchikov, Mechanisms of Thermal Quenching of Defect-Related Luminescence in Semiconductors, Phys. Status Solidi. 218 (2021) 2000101. https://doi.org/10.1002/pssa.202000101.

[59]   S. V. Nikiforov, V.S. Kortov, D.L. Savushkin, A.S. Vokhmintsev, I.A. Weinstein, Thermal quenching of luminescence in nanostructured monoclinic zirconium dioxide, Radiat. Meas. 106 (2017) 155–160. https://doi.org/10.1016/j.radmeas.2017.03.020.

[60]   A.M.A. Henaish, A.S. Vokhmintsev, I.A. Weinstein, Two-level quenching of photoluminescence in hexagonal boron nitride micropowder, AIP Conf. Proc. 1717 (2016) 040030. https://doi.org/10.1063/1.4943473.

[61]   S.S. Savchenko, A.S. Vokhmintsev, I.A. Weinstein, Non-radiative relaxation processes in luminescence of InP/ZnS quantum dots, J. Phys. Conf. Ser. 1537 (2020) 012015. https://doi.org/10.1088/1742-6596/1537/1/012015.

[62]   A.S. Vokhmintsev, I.A. Weinstein, Temperature effects in 3.9 eV photoluminescence of hexagonal boron nitride under band-to-band and subband excitation within 7–1100 K range, J. Lumin. 230 (2021) 117623. https://doi.org/10.1016/j.jlumin.2020.117623.

[63]   J. Zhang, J. Tolentino, E.R. Smith, J. Zhang, M.C. Beard, A.J. Nozik, M. Law, J.C. Johnson, Carrier transport in PbS and PbSe QD films measured by photoluminescence quenching, J. Phys. Chem. C. 118 (2014) 16228–16235. https://doi.org/10.1021/jp504240u.

[64]   S. Adam, D. V. Talapin, H. Borchert, A. Lobo, C. McGinley, A.R.B. De Castro, M. Haase, H. Weller, T. Möller, The effect of nanocrystal surface structure on the luminescence properties: Photoemission study of HF-etched InP nanocrystals, J. Chem. Phys. 123 (2005). https://doi.org/10.1063/1.2004901.

[65]   J. Laverdant, W.D. de Marcillac, C. Barthou, V.D. Chinh, C. Schwob, L. Coolen, P. Benalloul, P.T. Nga, A. Maître, Experimental Determination of the Fluorescence Quantum Yield of Semiconductor Nanocrystals, Materials (Basel). 4 (2011) 1182–1193. https://doi.org/10.3390/ma4071182.

[66]   M.G. Spirin, S.B. Brichkin, V.Y. Gak, V.F. Razumov, Influence of photoactivation on luminescent properties of colloidal InP@ZnS quantum dots, J. Lumin. 226 (2020) 117297. https://doi.org/10.1016/j.jlumin.2020.117297.

[67]   A.A. Guzelian, J.E.B. Katari, A. V. Kadavanich, U. Banin, K. Hamad, E. Juban, A.P. Alivisatos, R.H. Wolters, C.C. Arnold, J.R. Heath, Synthesis of size-selected, surface-passivated InP nanocrystals, J. Phys. Chem. 100 (1996) 7212–7219. https://doi.org/10.1021/jp953719f.

[68]    S. Sanguinetti, M. Henini, M. Grassi Alessi, M. Capizzi, P. Frigeri, S. Franchi, Carrier thermal escape and retrapping in self-assembled quantum dots, Phys. Rev. B. 60 (1999) 8276–8283. https://doi.org/10.1103/PhysRevB.60.8276.

[69]    G. Gélinas, A. Lanacer, R. Leonelli, R.A. Masut, P.J. Poole, Carrier thermal escape in families of InAs/InP self-assembled quantum dots, Phys. Rev. B. 81 (2010) 235426. https://doi.org/10.1103/PhysRevB.81.235426.

[70]    R.W. Collins, W. Paul, Model for the temperature dependence of photoluminescence in Si:H and related materials, Phys. Rev. B. 25 (1982) 5257–5262. https://doi.org/10.1103/PhysRevB.25.5257.

[71]    S. Kalytchuk, O. Zhovtiuk, S. V. Kershaw, R. Zbořil, A.L. Rogach, Temperature-Dependent Exciton and Trap-Related Photoluminescence of CdTe Quantum Dots Embedded in a NaCl Matrix: Implication in Thermometry, Small. 12 (2016) 466–476. https://doi.org/10.1002/smll.201501984.

[72]    I.A. Weinstein, V.S. Kortov, A.S. Vohmintsev, The compensation effect during luminescence of anion centers in aluminum oxide, J. Lumin. 122–123 (2007) 342–344. https://doi.org/10.1016/j.jlumin.2006.01.172.

[73]    A.F. Zatsepin, E.A. Buntov, A.L. Ageev, The relation between static disorder and photoluminescence quenching law in glasses: A numerical technique, J. Lumin. 130 (2010) 1721–1724. https://doi.org/10.1016/j.jlumin.2010.03.039.

[74]    V. Babin, K. Chernenko, L. Lipińska, E. Mihokova, M. Nikl, L.S. Schulman, T. Shalapska, A. Suchocki, S. Zazubovich, Y. Zhydachevskii, Luminescence and excited state dynamics of $Bi^{3+}$ centers in $Y_2O_3$, J. Lumin. 167 (2015) 268–277. https://doi.org/10.1016/j.jlumin.2015.06.029.

[75]    I. V. Ignatiev, I.E. Kozin, S. V. Nair, H.W. Ren, S. Sugou, Y. Masumoto, Carrier relaxation dynamics in InP quantum dots studied by artificial control of nonradiative losses, Phys. Rev. B - Condens. Matter Mater. Phys. 61 (2000) 15633–15636. https://doi.org/10.1103/PhysRevB.61.15633.

[76]    S. Adachi, Properties of Group-IV, III-V and II-VI Semiconductors, John Wiley & Sons, Ltd, Chichester, UK, 2005. https://doi.org/10.1002/0470090340.

[77]    J. Jasieniak, M. Califano, S.E. Watkins, Size-dependent valence and conduction band-edge energies of semiconductor nanocrystals, ACS Nano. 5 (2011) 5888–5902. https://doi.org/10.1021/nn201681s.

[78]    P.E. Lippens, M. Lannoo, Calculation of the band gap for small CdS and ZnS crystallites, Phys. Rev. B. 39 (1989) 10935–10942. https://doi.org/10.1103/PhysRevB.39.10935.

[79]    E.M. Chan, C. Xu, A.W. Mao, G. Han, J.S. Owen, B.E. Cohen, D.J. Milliron, Reproducible, high-throughput synthesis of colloidal nanocrystals for optimization in multidimensional parameter space, Nano Lett. 10 (2010) 1874–1885. https://doi.org/10.1021/nl100669s.